\documentstyle[twoside,fleqn,espcrc2]{article}
\let\a \alpha    \let\b \beta     \let\g \gamma    
\let\d \delta
\let\e \epsilon
      \let\h \eta      \let\q \theta

\let\l \lambda                    
\let\x \xi

        \let\s \sigma              

\let\f \phi
                   
\let\vf \varphi
\let\w \omega

\let\la \label    \let\nn \nonumber

      \let\del \partial      \let\bm \bibitem

\newcommand{\be}{\begin{equation}}
\newcommand{\ee}[1]{\label{#1}\end{equation}}
\newcommand{\nee}{\nonumber\end{equation}}
\newcommand{\bea}{\begin{eqnarray}}
\newcommand{\eea}{\end{eqnarray}}
\newcommand{\ra}{\rightarrow}

\newcommand{\NPB}[3]{Nucl. Phys. B#1 (#2) #3}
\newcommand{\CMP}[3]{Commun. Math.\ Phys.\ #1 (#2) #3}
\newcommand{\PRD}[3]{Phys. Rev.\ D#1 (#2) #3}
\newcommand{\PLB}[3]{Phys. Lett.\ B#1 (#2) #3}
\newcommand{\tmath}[1]{\mbox{$#1$}}

\newcommand{\refer}[1]{(\ref{#1})}

\newcommand{\qb}{\bar{\q}}
\newcommand{\lp}{{\l^\prime}}
\newcommand{\lpp}{{\l^{\prime\prime}}}
\newcommand{\lppp}{{\l^{\prime\prime\prime}}}
\newcommand{\whQ}{{\widehat{Q}}}
\newcommand{\ft}[2]{{\textstyle\frac{#1}{#2}}}

\newcommand{\AmS}{{\protect\the\textfont2
  A\kern-.1667em\lower.5ex\hbox{M}\kern-.125emS}}

\title{The Supermembrane with Winding}
\author{Bernard de Wit\address{Institute for Theoretical Physics, 
Utrecht University, Princetonplein 5, 3508 TA Utrecht,\\ The Netherlands}
, Kasper Peeters\address{NIKHEF, P.O. Box 41882, 1009 DB Amsterdam, 
 The Netherlands} and Jan~C.\ Plef\/ka$^{\mbox{\scriptsize b}}$}
\begin{document}
\setlength{\arraycolsep}{0pt}
\begin{titlepage}
\begin{flushright} THU-97/20\\ 
 NIKHEF 97-032\\[2mm] 
{\tt hep-th/9707261}
\end{flushright}
\vfill
\begin{center}
{\large\bf The Supermembrane with Winding${}^\dagger$}\\
\vskip 7.mm
{Bernard de Wit${}^{\mbox{\scriptsize a}}$, Kasper 
Peeters${}^{\mbox{\scriptsize b}}$ and Jan~C.\  
Plef\/ka${}^{\mbox{\scriptsize b}}$}\\
\vskip 0.2cm                                                      
{${}^{\mbox{\scriptsize a}}$\em Institute for Theoretical Physics, 
Utrecht University}\\ 
{\em Princetonplein 5, 3508 TA Utrecht, The Netherlands} \\
\vskip .1 cm
{${}^{\mbox{\scriptsize b}}$\em NIKHEF, P.O. Box 41882}\\
{\em  1009 DB Amsterdam,  The Netherlands} 
\end{center}
\vfill
\begin{center}
{\bf ABSTRACT}
\end{center}
\begin{quote}
The supersymmetry algebra for supermembranes, quantized in the 
light-cone gauge, exhibits central 
charges induced by wrapping the membrane around compact 
dimensions. These central charges are manifestly consistent with Lorentz 
symmetry. While the central charges raise the mass of the 
membrane states, they still leave the mass spectrum continuous, 
at least generically. The lower bound on the mass spectrum is set 
by the winding number and corresponds to a BPS state.
\vfill      \hrule width 5.cm
\vskip 2.mm
{\small\small
\noindent $^\dagger$ Invited talk given at SUSY '97 by 
B.~d.W., Philadelphia, PA, USA, 27-31 May 1997; to appear in 
Nuclear Physics B (Proc. Suppl.).}   
\end{quote}
\begin{flushleft}
July 1997
\end{flushleft}
\end{titlepage}
\begin{abstract}
The supersymmetry algebra for supermembranes, quantized in the 
light-cone gauge, exhibits central 
charges induced by wrapping the membrane around compact 
dimensions. These central charges are manifestly consistent with Lorentz 
symmetry. While the central charges raise the mass of the 
membrane states, they still leave the mass spectrum continuous, 
at least generically. The lower bound on the mass spectrum is set 
by the winding number and corresponds to a BPS state.
\end{abstract}                                          

\maketitle

Supersymmetric matrix models that correspond to the reduction of 
supersymmetric gauge theories to one (time) dimension \cite{CH}, are 
relevant for a variety of problems. With an 
infinite-dimensional gauge group consisting of the  
area-preserving diffeomorphisms of a two-dimensional surface of 
certain topology, the model describes the quantum mechanics of a
supermembrane \cite{bergs87} in the light-cone gauge \cite{BSTa,dWHN}. 
The group of area-preserving diffeomorphisms (or, at 
least, its simple subgroup) can be approximated by SU($N$) in 
the large-$N$ limit \cite{GoldstoneHoppe,dWHN}. More recently, it was 
discovered that the collective dynamics of D-branes 
\cite{Dbranes} is also described by various dimensional 
reductions of supersymmetric  U($N$) gauge 
theories, where $N$ now corresponds to the number of branes 
\cite{boundst}. Supersymmetric U($N$) quantum mechanics thus describes the 
dynamics of $N$ D-particles \cite{Dpart} and represents a 
regularized version of the light-cone supermembrane. 

Here we report some recent results on supermembranes 
that are wrapped around compact dimensions in a flat 
eleven-dimensional spacetime \cite{DWPP}. We consider the supersymmetry 
algebra and show that it remains consistent with 
eleven-dimensional Lorentz invariance. Furthermore we present a 
qualitative discussion on the nature of 
the mass spectrum for wrapped membranes and present arguments 
as to why it will remain continuous. These results are derived in 
the context of   
the supersymmetric quantum mechanics that is obtained in the 
light-cone gauge. We start by introducing the supersymmetric gauge
theory of area-preserving diffeomorphisms, paying 
attention to various issues that are relevant for the 
supermembrane winding.  

The area-preserving diffeomorphisms on the membrane spacesheet, 
parametrized by $\sigma^1, \sigma^2$, are defined by 
\tmath{\s^r \ra \s^r + \x^r(\s)} with 
\be
\del_r(\sqrt{w(\s)}\, \x^r(\s)\, )=0,
\ee{APD}
where \tmath{w(\sigma )} is the spacesheet density, whose integral is 
normalized to unity. We may rewrite this condition in terms of 
dual spacesheet vectors by  
\be                                  
\sqrt{w(\s)}\,\x^r(\s)= \e^{rs}\, F_s(\s)\, .
\ee{1form}
In the language of differential forms the
condition \refer{APD} is then be simply recast as \tmath{{\rm 
d}F=0}. The trivial solutions are the exact forms \tmath{F={\rm d}\x}, 
or, in components, 
\be
F_s=\del_s\x(\s),
\ee{exact}
for any globally defined function $\x(\s)$. The nontrivial 
solutions are the closed forms that are not exact. On a Riemann surface of
genus $g$ there are precisely $2g$ linearly independent non-exact 
closed forms, whose integrals along the homology cycles are 
normalized to unity.\footnote{%
  In the mathematical literature the globally defined exact forms 
  are called ``hamiltonian vector fields'', whereas the closed 
  but not exact forms which are not globally defined go under the 
  name ``locally hamiltonian vector fields''.} %
In components we write
\be
F_s=\f_{(\l)\, s} \qquad \l=1,\ldots,2g.
\ee{harm}

Here we note that the exact
vectors generate an invariant subgroup
\cite{dWMN}. This follows from the commutator of two
infinitesimal transformations corresponding to the forms $F^{(1)}_r$ and
$F^{(2)}_r$, which yields an infinitesimal transformation
defined by
\be
F^{(3)}_r=\del_r\, \bigg(\frac{\e^{st}}{\sqrt{w}}\, F^{(2)}_s\, 
F^{(1)}_t\, \bigg).
\ee{comm}
The presence of the closed but non-exact forms is crucial 
for the winding of the embedding membrane coordinates.

In the light-cone formulation the Hamiltonian for a supermembrane 
moving in a flat eleven-dimensional spacetime, takes the form 
\bea
{\cal H}&=& \frac{1}{P_0^+}\, \int {\rm d}^2\s \, \sqrt{w}\,
\bigg[ \, \frac{P^a\, P_a }{2\,w} + \ft{1}{4} \{\, 
X^a,X^b\,\}^2 \nn \\
&&\hspace{25mm}
-P^+_0\, \qb\,\g_- \g_a\, \{\, X^a , \q\,\}\, \bigg]\, .\la{memham}
\eea
Here the integral runs over the spatial components of the
worldvolume denoted by $\s^1$ and $\s^2$.
In the above  $X^a(\s)$ ($a=1,\ldots,9$) denote the transverse 
target--space embedding coordinates lying in $T^d\times{\bf R}^{9-d}$ and 
thus permitting us to have winding on the $d$-dimensional torus $T^d$. 
Accordingly $P^a(\s)$ are their conjugate  momenta. In this gauge 
the light-cone coordinate $X^+$ is linearly related to the 
world-volume time. The momentum $P^+$ is time independent and 
proportional to the center-of-mass value $P^+_0$ times the spacesheet
density ${\sqrt{w(\s)}}$. The center-of-mass momentum
$P_0^-$ is equal to minus the Hamiltonian \refer{memham}.
Moreover we have the fermionic variables $\q(\s)$, which are
32-component Majorana spinors subject to the gauge 
condition \tmath{\g_+\, \q=0}. And finally we made use of the
Poisson bracket \tmath{\{ A,B\} } defined by
\be
\{ A(\s ),B(\s )\} = \frac{1}{\sqrt{w(\s)}}\, \e^{rs}\, \del_r A(\s )\,
\del_s B(\s ).
\ee{poisbrak}
Note that the coordinate $X^-$ itself does
not appear in the Hamiltonian \refer{memham}. It is defined by 
\be
P^+_0\, \del_rX^-= - \frac{{\bf P} \cdot \del_r{\bf X}}{\sqrt{w}} - 
P^+_0\, \qb\g_-\del_r\q\,,
\ee{delxminus}
and implies a number of constraints that will be important in 
the following.

Whereas the momenta ${\bf P}(\s)$ and the fermionic coordinates 
$\theta(\s)$ remain single valued on the spacesheet, the 
embedding coordinates, written as one-forms with components 
$\del_r {\bf X}(\s)$ and  $\del_r X^-(\s)$, are decomposed into 
closed forms. Their non-exact contributions are multiplied by an 
integer times the length of the compact direction.
The constraint alluded to above,  
amounts to the condition that the right-hand side of 
\refer{delxminus} is closed. 

The light-cone supermembrane and the collective dynamics of D-branes
 are related in the context of M-theory 
\cite{Mtheory}, the conjectured eleven-dimensional theory, which, upon 
compactification of the eleventh dimension to a circle, yields 
type-IIA string theory. {}From a ten-dimensional viewpoint
the Kaluza-Klein states emerging from the compactification of 
eleven-dimensional supergravity 
carry a Ramond-Ramond charge, just as generic D-branes 
\cite{Polchinski}. The Kaluza-Klein states with lowest 
nonzero charge can thus be identified with the D-particles, 
which, from the perspective of IIA supergravity, can be identified as 
extremal black holes \cite{Townsend,witten3}. 
The supermembrane can presumably be viewed  
as the result of the collective dynamics of arbitrarily 
large numbers of D-particles. This would then naturally explain 
the continuum of the supermembrane mass spectrum \cite{dWLN}. 

These and other considerations led to the proposal that the degrees of
freedom of M-theory are in fact captured in the large-$N$ limit of
supersymmetric matrix models \cite{BFSS}. 
Viewed as a theory of 
D-particles it has been shown that the long-distance 
interactions between two particles is consistent with 
eleven-dimensional supergravity \cite{Dpart,BFSS,BBPT}.
By T-duality arguments one can define
compactified versions of the supersymmetric matrix models
\cite{BFSS,Tdual} and incorporate the effect of virtual winding strings
stretched between the D-particles.  
Recently further 
light has been shed on the relation between the matrix models and 
the known string theories \cite{matrixstrings}, which involves 
second-quantized string states. The appearance of a second-quantized
string spectrum seems to fit in with the instability of the
supermembrane, which collapses into states of multi-membranes  
connected by infinitely thin strings of arbitrary length. In the
context of the matrix model this phenomenon was stressed already in
\cite{BFSS}.

One of the crucial aspects of the proposal of \cite{BFSS} is that the
supersymmetric matrix models lack Lorentz invariance. 
It is known that Lorentz invariance is realized in the
$N\to\infty$ limit (at least, classically), although one needs to
specify additional data, which can be extracted from the supermembrane
\cite{dWMN,EMM}. We therefore wish to focus on supermembranes in
flat, compact target spaces, yielding a possibly different, but 
well-defined, viewpoint on the compactified large-$N$ models. 
Unfortunately we were unable to find a suitable generalization of 
the SU($N$) supersymmetric matrix model regularization of the 
supermembrane to the winding case. 

Let us now return to the issue of defining a supersymmetric gauge 
theory of area--preserving diffeomorphisms. While it has been 
known for quite some time \cite{dWHN} that the   
light--cone supermembrane can be formulated in terms of such a 
gauge theory, emphasizing the membrane's residual
gauge symmetry from the start, it is a priori not obvious 
whether this equivalence continues to hold after introducing 
winding contributions.  
Under the full group of area--preserving diffeomorphisms the fields $X^a$,
$X^-$ and $\q$ transform according to
\bea
\d X^a&=& {\e^{rs}\over \sqrt{w}}\, \x_r\, \del_s X^a\,,
\quad 
\d X^-= {\e^{rs}\over \sqrt{w}}\, \x_r\, \del_s X^-\,,
\nn \\
\d \q^a&=& {\e^{rs}\over \sqrt{w}}\, \x_r\, \del_s \q\,,
\la{APDtrafoXtheta}
\eea
where the time--dependent reparametrization $\x_r$ consists of
closed exact and non-exact parts. Accordingly there is a gauge
field $\w_r$, which is therefore closed as well, transforming as
\be
\d\w_r=\del_0\x_r + \del_r \bigg( {\e^{st}\over\sqrt{w}}\,\x_s\,\w_t\bigg), 
\ee{APDtrafoomega}
and corresponding covariant derivatives
\bea
D_0 X^a&=& \del_0X^a - {\e^{rs}\over \sqrt{w}}\, \w_r\, \del_s X^a, 
\nn \\
D_0 \q &=& \del_0\q - {\e^{rs}\over \sqrt{w}}\, \w_r\, \del_s\q,
\la{covderiv}
\eea
and similarly for \tmath{D_0 X^-}. 

The action corresponding to the following lagrangian density is
then gauge invariant under the 
transformations \refer{APDtrafoXtheta} and \refer{APDtrafoomega},
\bea
{\cal L}&=&P^+_0\,\sqrt{w}\, \Big[\,  
\ft{1}{2}\,(D_0{\bf X})^2 + \qb\,\g_-\,
D_0\q  \la{gtlagrangian}    \\
&&\hspace{14mm} - \ft{1}{4}\,(P^+_0)^{-2}\,  \{ X^a,X^b\}^2 
\nn\\
&&\hspace{14mm} + (P^+_0)^{-1}\, \qb\,\g_-\,\g_a\,\{X^a,\q\} +
  D_0 X^-\Big]\, , \nn
\eea
where we draw attention to the last term proportional to
$X^-$, which can be dropped in the absence of winding and did not
appear in \cite{dWHN}. The action corresponding to
\refer{gtlagrangian} is also invariant under the 
supersymmetry transformations  
\bea
\d X^a &=& -2\, \bar{\e}\, \g^a\, \q\,, \nn\\
\d \q  &=& \ft{1}{2} \g_+\, (D_0 X^a\, \g_a + \g_- )\, \e
\nn\\&& +\ft{1}{4}(P^+_0)^{-1} \,
\{ X^a,X^b \}\, \g_+\, \g_{ab}\, \e ,\nn\\
\d \w_r &=& -2\,(P^+_0)^{-1}\, \bar{\e}\,\del_r\q\,.
\la{susytrafos}
\eea
The supersymmetry variation of $X^-$ is not relevant and may be
set to zero. 
The full equivalence with the membrane Hamiltonian is now established by
choosing the \tmath{\w_r=0} gauge and passing to the Hamiltonian 
formalism. The field equations for $\w_r$ then lead to
the membrane constraint \refer{delxminus} (up to exact contributions), 
partially defining \tmath{X^-}.
Moreover the Hamiltonian corresponding to the gauge theory lagrangian of 
\refer{gtlagrangian} is nothing
but the light--cone supermembrane Hamiltonian \refer{memham}.
Observe that in the above gauge-theoretical construction, the 
space-sheet metric $w_{rs}$ enters only through its density 
$\sqrt{w}$ and hence 
vanishing or singular metric components do not pose problems.

We are now in a position to study the full eleven--dimensional 
supersymmetry algebra of the winding supermembrane. For this we 
decompose the supersymmetry charge $Q$ associated with the 
transformations \refer{susytrafos} as \tmath{Q= Q^+ + Q^-}, where
\be
 Q^\pm = \ft{1}{2}\, \g_\pm\,\g_\mp\, Q, 
\ee{Qdecomposition}
to obtain
\bea
Q^+&=&\int {\rm d}^2 \s \, \Big(\, 2\, P^a\, \g_a + \sqrt{w}\, \{\,
X^a, X^b\, \} \, \g_{ab}\, \Big) \, \q , \nn \\
Q^-&=& 2\, P^+_0\, \int {\rm d}^2\s\, \sqrt{w}\, \g_-\, \q .
\la{Q-cont}
\eea
The canonical Dirac brackets are derived by the standard
methods and read
\bea
(\, X^a(\s), P^b(\s^\prime)\, )_{\mbox{\tiny DB}} &=& \d^{ab}\, 
\d^2(\s-\s^\prime)\, , \\
(\, \q_\a(\s), \qb_\b(\s^\prime)\, )_{\mbox{\tiny DB}} &=&
\ft{1}{4}\,\frac{(\g_+)_{\a\b}\,}{P^+_0 \,w^{1/2}}
\, \d^2(\s-\s^\prime). \nn
\eea
In the presence of winding the results given in \cite{dWHN} yield the
supersymmetry algebra
\bea
&&(\, Q^+_\a, \bar{Q}^+_\b\, )_{\mbox{\tiny DB}} =
2\, (\g_+)_{\a\b}\, {\cal H} \nn\\ 
&& \hspace{8mm} - 2\,  (\g_a\, \g_+)_{\a\b}\, \int
{\rm d}^2\s\, \sqrt{w}\, \{\, X^a, X^-\,\}\, , \nn\\
&&(\, Q^+_\a, \bar{Q}^-_\b\, )_{\mbox{\tiny DB}} =
-(\g_a\,\g_+\,\g_- )_{\a\b} \, P^a_0 \nn\\
&&\hspace{8mm} - \ft{1}{2}\,(\g_{ab}\, \g_+\g_- )_{\a\b}\,\int {\rm
d}^2\s\,\sqrt{w}\, \{\, X^a,X^b\,\}\, , \nn\\
&&(\, Q^-_\a, \bar{Q}^-_\b\, )_{\mbox{\tiny DB}} 
= -2\, (\g_- )_{\a\b}\, P^+_0 \, , \la{contsusy}
\eea
where use has been made of the defining equation
\refer{delxminus} for $X^-$. 
The new feature of this supersymmetry algebra is the emergence of the 
central charges in the first two anticommutators, which are
generated through the winding contributions.
They represent topologically conserved quantities obtained by integrating
the winding densities \tmath{z^{a}(\s)=\e^{rs}\,\del_r X^a\,\del_s X^-} 
and \tmath{z^{ab}(\s) =\e^{rs}\,\del_r X^a\,\del_s X^b} over the
space-sheet. It is gratifying to observe the manifest
Lorentz invariance of \refer{contsusy}. Here  we should point out
that, in adopting the light-cone gauge, we assumed that there was
no winding for the coordinate $X^+$. Hence the
consistent picture emerges, that supermembranes may
live in spacetimes subject to the sourcefree equations of 
motion of eleven dimensional supergravity, with central charges 
induced by their own winding.
In \cite{BSS} the corresponding algebra for the matrix 
regularization was studied. It coincides with ours in the 
large-$N$ limit, in which an additional longitudinal five-brane
charge vanishes, provided that one identifies the longitudinal
two-brane charge with the central charge in the
first line of \refer{contsusy}. This requires the definition of
$X^-$ in the matrix regularization, a topic that was dealt with in
\cite{dWMN}. We observe that the discrepancy noted in \cite{BSS}
between the matrix calculation and 
certain surface terms derived in \cite{dWHN}, seems to have no
consequences for the supersymmetry algebra. A possible reason for
this could be that certain Schwinger terms have not been treated
correctly in the matrix computation, as was claimed in a recent
paper \cite{EMM}. 

In order to define a matrix approximation one introduces a complete 
orthonormal basis of functions $Y_A(\s)$ for the globally defined
$\x(\s)$ of \refer{exact}. One may then write down the following
mode expansions for the phase space variables of the
supermembrane, 
\bea
\del_r{\bf X}(\s) &=& {\bf X}^\l\, \f_{(\l)\, r} + \sum_A {\bf X}^A\, 
\del_r Y_A(\s),\nn \\
{\bf P}(\s) &=& \sum_A \sqrt{w}\, {\bf P}^A\, Y_a(\s), \nn\\
\q(\s) &=& \sum_A \q^A\, Y_A(\s) , \la{modeexp}
\eea
introducing winding modes for the transverse coordinates $X^a$. A 
similar expansion exists for $X^-$.  One then naturally
introduces the structure constants of the group of area--preserving 
diffeomorphism by \cite{dWMN}
\bea
f_{ABC} &=& \int {\rm d}^2\s\, \e^{rs}\,\del_r Y_A\, \del_s Y_B\,
Y_C\,, \nn \\ 
f_{\l BC} &=& \int {\rm d}^2\s\, \e^{rs}\,\f_{(\l)\, r}\, \del_s
Y_B\, Y_C\,, \nn \\ 
f_{\l \lp C} &=& \int {\rm d}^2\s\, \e^{rs}\,\f_{(\l)\, r}\,
\f_{(\lp)\, s}\, Y_C \, . 
\eea
Note that with $Y_0=1$, we have \tmath{f_{AB0}=f_{\l B0}=0}.
The raising and lowering of the $A$ indices is performed with the 
invariant metric
\tmath{
\h_{AB}= \int {\rm d}^2\s\, \sqrt{w}\, Y_A(\s)\, Y_B(\s)
}
and there is no need to introduce a metric for the $\l$ indices.

By plugging the mode expansions \refer{modeexp} into the Hamiltonian 
\refer{memham} one obtains the decomposition
\bea
{\cal H} &=&
 \ft{1}{2}\, {\bf P}_0\cdot{\bf P}_{0}+ \ft{1}{2}\, {\bf 
P}^{A}\cdot {\bf P}_{A} 
\la{memhammode}\\
&&
 - f_{ABC}\, \qb^C\, \g_-\,\g_a\,
\q^B\, X^{a\, A} 
\nn\\&& - f_{\l BC}\, \qb^C\, \g_-\,\g_a\,\q^B\, X^{a\, \l} 
\nn\\ &&+ \ft{1}{4}\, 
{f_{\l \lp}}^0\, f_{\lpp \lppp 0}\, X^{a\, \l}\, X^{b\, \lp}\, 
X^\lpp_a\, X^\lppp_b 
\nn\\&&
 + \ft{1}{4}\, {f_{AB}}^E\, f_{CDE}\, X^{a\, A}\, X^{b\, B}\, X^C_a\, X^D_b
\nn\\&&
+ {f_{\l B}}^E\, f_{CDE}\, X^{a\, \l}\, X^{b\, B}\, X^C_a\, X^D_b 
\nn\\&&
+\ft{1}{2}\,  {f_{\l B}}^E\, f_{\lp DE}\, X^{a\, \l}\, X^{b\, B}\, 
X^\lp_a\, X^D_b 
\nn\\&&
+\ft{1}{2}\,  {f_{\l B}}^E\, f_{C \lp E}\, X^{a\, \l}\, 
X^{b\, B}\, X^C_a\, X^\lp_b 
\nn\\&&
+ \ft{1}{2}\,  {f_{\l \lp}}^E\, f_{C DE}\, X^{a\, \l}\, X^{b\, \lp}\, 
X^C_a\, X^D_b 
\nn\\&&
+ {f_{\l \lp}}^E\, f_{\lpp DE}\, X^{a\, \l}\, X^{b\, \lp}\, 
X^\lpp_a\, X^D_b 
\nn\\&&
+ \ft{1}{4}\, {f_{\l \lp}}^E\, f_{\lpp \lppp E}\, X^{a\, \l}\, X^{b\, \lp}
\, X^\lpp_a\, X^\lppp_b,
\nn
\eea
where here and henceforth we spell out the zero-mode dependence
explicitly, i.e.\ the range of values for $A$ no longer includes
$A=0$. Note that for the toroidal supermembrane 
\tmath{f_{\l\lp A}=0} and thus the last three terms in \refer{memhammode}
vanish. The fourth term 
represents the winding number squared. In the matrix formulation,
the winding number takes the form of a trace over a commutator.
We have scaled the Hamiltonian by a 
factor of $P^+_0$ and  the fermionic variables by a factor $(P^+_0)^{-1/2}$.
Supercharges will be rescaled as well, such as to eliminate
explicit factors of $P^+_0$. 

The constraint equation \refer{delxminus} is translated into mode
language by contracting it with \tmath{\e^{rs}\, \f_{(\l)\, s}} and
\tmath{\e^{rs}\, \del_s Y_C}, respectively, and integrating the result
over the spacesheet to obtain the two constraints, 
\bea
\vf_\l &=& f_{\l\lp 0}( {\bf X}^\lp\cdot{\bf P}_0 +X^{- 
\lp} P^+_0 ) 
+ f_{\l\lp C}\, {\bf X}^\lp\cdot{\bf P}^C \nn\\
&& + f_{\l BC}\,(\,  {\bf X}^B\cdot{\bf P}^C +\qb^C\,\g_-\, \q^B\, )
=0,\la{q1} \\
\vf_A &=& f_{ABC}\, (\, {\bf X}^B\cdot{\bf P}^C + \qb^C\,\g_-\,\q^B\, )
\nn\\ 
&& + f_{A\l C} \, {\bf X}^\l\cdot{\bf P}^C =0  \, , \la{q2}
\eea
taking also possible winding in the $X^-$ direction into account.
Note that even for the non--winding case \tmath{X^{a\,\l}=0}, 
there remain the extra $\vf_\l$ constraints.

The zero-mode contributions completely decouple in the
Hamiltonian and the supercharges. We thus perform a split in
$Q^+$ treating zero modes and fluctuations separately to obtain
the mode expansions,  
\be
Q^- = 2\,\g_-\, \q^0 \,, \qquad Q^+ = Q^+_{(0)} + \whQ^+\,, 
\ee{split}
where
\bea
Q^+_{(0)} &=&  \Bigl (\,2\, P^a_0\,\g_a + f_{\l\lp 0}\,
X^{a\,\l}\, X^{b\,\lp}\, \g_{ab}\, \Bigr )\, \q_0 \,, \nn \\
\widehat{Q}^+ &=& \Bigl (\, 2\, P^a_C\,\g_a + 
f_{ABC}\, X^{a\, A}\, X^{b\, B}\,\g_{ab} \nn\\
&&\hspace{2mm} + 2\, f_{\l BC}\, X^{a\, \l}\, X^{b\, B}\,\g_{ab}\nn\\
&& \hspace{2mm} + f_{\l\lp
C}\, X^{a\, \l}\, X^{b\, \lp}\,\g_{ab}\,  
\Bigr )\, \q^C \,.
\eea
Upon introducing the supermembrane mass operator by
\be 
{\cal M}^2=2\, {\cal H} - {\bf P}_0\cdot{\bf P}_{0}- \ft{1}{2}\,(
f_{\l\lp 0}\, X^{a\, \l}\, X^{b\, \lp})^2 ,
\ee{mass2}
the supersymmetry algebra \refer{contsusy} then takes the form
\bea
&&\{\whQ{}^+_\a, \bar{\widehat{Q}}{}^+_\b \} = (\g_+)_{\a\b} \, 
{\cal M}^2 \label{s1}\\[.5mm]
&&\hspace{4mm} -2 (\g_a\g_+ )_{\a\b}\, f_{\l\lp 0}\, X^{a\,\l} ( X^{- 
\lp}\, P^+_0 +  {\bf X}^\lp\cdot {\bf P}_0 ) \,, \nn\\[2mm]
&&\{ Q^+_{(0)\,\a }, \bar{Q}^+_{(0)\,\b }\, \} =  \label{w3} \\
&&\hspace{12mm} (\g_+)_{\a\b}\,\Big( {\bf P}_0\cdot{\bf P}_{0} 
+\ft{1}{2} 
(f_{\l\lp 0}\,X^{a\, \l}\, X^{b\, \lp})^2\Big) \nn\\
&& \hspace{12mm}
 +2\, (\g_a\,\g_+ )_{\a\b}\, f_{\l\lp 0}\, X^{a\,\l}\, {\bf X}^\lp\cdot 
{\bf P}_0  \,, \nn\\[1.5mm]
&& \{ Q^+_{(0)\,\a } , \bar{Q}^-_\b\, \} = -(\g_a\,\g_+\,\g_-)_{\a\b}\,
P^a_0  \label{s4}\\
&& \hspace{12mm}- \ft{1}{2}\, (\g_{ab}\,\g_+\, \g_-)_{\a\b}\, 
f_{\l\lp 0}\, X^{a\, \l}\, X^{b\, \lp }\, , \nn\\[.5mm]
&& \{ \whQ^+_\a, \bar{Q}^-_\b\, \} = \{\, Q^+_{(0)\,\a } \, , 
\bar{\widehat{Q}}{}^+_\b\, \} = 0 \,. 
\la{modesusy}
\eea
And the mass operator commutes with all the supersymmetry charges,
\be
[\, \whQ^+, {\cal M}^2\, ] =[\, Q_{(0)}^+, {\cal M}^2\, ] =
[\, Q^-, {\cal M}^2\, ] = 0 ,
\ee{QswithM2}
defining a supersymmetric quantum-mechanical model. 

At this stage it would be desirable to present a matrix
model regularization of the supermembrane with winding contributions,
generalizing the SU($N$) approximation to the exact subgroup of
area-preserving diffeomorphisms \cite{GoldstoneHoppe,dWHN}, at
least for toroidal 
geometries. However, this program seems to fail due to the fact that the 
finite-$N$ approximation to the structure constants \tmath{f_{\l BC}} 
violates the Jacobi identity, as was already noticed in \cite{dWMN}. 

Finally we turn to the question of the mass spectrum for 
membrane states with winding. The mass spectrum 
of the supermembrane without winding is continuous. This was 
proven in the SU($N$) regularization \cite{dWLN}. Whether or 
not nontrivial zero-mass states exist, is not rigorously known 
(for some discussion on these questions, we refer the reader to 
\cite{DWN,HF,SS}). Those would coincide with the states of
eleven-dimensional supergravity. It is often   
argued that the winding may remove the continuity of the 
spectrum (see, for instance, \cite{Russo}). 
There is no question that winding may increase the 
energy of the membrane states. A membrane winding around more 
than one compact  
dimension gives rise to a nonzero central charge in the 
supersymmetry algebra. This central charge sets a lower limit on  
the membrane mass. 
However, this should not be interpreted as an indication that 
the spectrum becomes discrete. The possible continuity of the spectrum 
hinges on two features. First the system should possess 
continuous valleys of classically degenerate states. 
Qualitatively one recognizes immediately that this feature is not 
directly affected by the winding. A classical membrane with 
winding can still have stringlike configurations of arbitrary length,  
without increasing its area. Hence the classical instability 
still persists. 

The second feature is supersymmetry. Generically the classical 
valley structure is lifted by quantum-mechanical corrections, 
so that the wave function cannot escape to infinity. This 
phenomenon can be 
understood on the basis of the uncertainty principle. Because, at 
large distances, the valleys become increasingly narrow, the wave 
function will be squeezed more and more which tends to induce an 
increasing spread in its momentum. This results in an increase of 
the kinetic energy. Another way to understand this is by noting 
that the transverse oscillations perpendicular to the valleys 
give rise to a zero-point energy, which acts as an effective 
potential barrier that confines the wave function. When the 
valley configurations are supersymmetric the contributions from 
the bosonic and the fermionic transverse oscillations cancel each 
other, so that the wave function will not be confined and can
extend arbitrarily far into the valley. This phenomenon indicates
that the energy spectrum must be continuous. 

Without winding it is clear that the valley configurations are 
supersymmetric, so that one concludes that the spectrum is 
continuous. With winding the latter aspect is somewhat more 
subtle. However, we note that, when the winding density is
concentrated in one part of the spacesheet, then valleys can
emerge elsewhere
corresponding to stringlike configurations with supersymmetry.
Hence, as a space-sheet {\it local} field theory,  
supersymmetry can be broken in one region where the winding is 
concentrated and unbroken in another. In the latter region 
stringlike configurations can form,  
which, at least semiclassically, will not be suppressed by 
quantum corrections. 

We must stress that we are describing only the 
generic features of the spectrum. Our arguments by no means 
preclude the existence of mass gaps. To prove or disprove the 
existence of  
discrete states is extremely difficult. While the contribution of 
the bosonic part of the Hamiltonian increases by concentrating the 
winding density on part of the spacesheet, the matrix elements in 
the fermionic directions will also grow large, making it 
difficult to estimate the eigenvalues. At this moment the only 
rigorous result is the BPS bound that follows from the supersymmetry 
algebra. Obviously, the state of  lowest mass for 
given winding numbers, is always a BPS state, which is invariant under some
residual supersymmetry.

Meanwhile a number of papers have appeared that also address some of the 
issues discussed above \cite{recent}.

\bigskip\noindent
\leftline{\bf Acknowledgements:} 
B.~d.W. thanks the organizers of SUSY '97 for their kind 
hospitality. He also thanks T. Banks and J. Maldacena for 
discussions. This work was supported by the 
European Commission HCM program ERBCHR-CT92-0035.



\begin{thebibliography}{9}
\bibitem{CH} M.\ Claudson and M.B.\ Halpern, Nucl.\ Phys.\ B250 (1985) 
689;\\ 
R.\ Flume, Ann.\ Phys.\ {164} (1985) 189; \\
M.\ Baake, P.\ Reinicke, and V.\ Rittenberg, J.\ Math.\
Phys.\ {26} (1985) 1070. 
\bm{bergs87} E.A.\ Bergshoeff, E.\ Sezgin and P.K.\ Townsend, 
\PLB{189}{1987}{75}.
\bibitem{BSTa} E.A.\ Bergshoeff, E.\ Sezgin and Y.\ Tanii, Nucl.\ 
Phys.\ B298 (1988) 187.
\bm{dWHN} B.\ de Wit, J.\ Hoppe and H.\ Nicolai, \NPB{305}{1988}{545}.
\bibitem{GoldstoneHoppe} J.\ Goldstone, unpublished; J.\ Hoppe, in 
proc.\ Int.\ Workshop on 
Constraint's Theory and Relativistic Dynamics, eds.\ G.\ Longhi and
L.\ Lusanna (World Scient., 1987).
\bibitem{Dbranes} J.\ Polchinski, S.\ Chaudhuri and C.V.\ Johnson, 
{\it Notes on D-branes}, {\tt hep-th/9602052}; \\
J.\ Polchinski, {\it TASI 
Lectures on D-branes}, {\tt hep-th/9611050};\\ 
C.\ Bachas, {\it (Half) A 
Lecture on D-branes}, lectures given at the Workshop on Gauge 
Theories, Applied Supersymmetry and Quantum Gravity, London, 1996, 
{\tt hep-th/9701019}.
\bibitem{boundst} E. Witten, Nucl.\ Phys.\ B460 (1996) 335, {\tt
hep-th/9510135}. 
\bibitem{Dpart}
U.\ Danielson, G. Ferretti and B. Sundborg, Int. J. Mod. Phys. A11 
(1996) 5463, {\tt hep-th/9603081};\\ 
D. Kabat and P. Pouliot, Phys. Rev. Lett. 77 (1996) 1004, 
{\tt hep-th/9603127}; \\
M.R.\ Douglas, D.\ Kabat, P.\ Pouliot and S.H.\ Shenker, 
Nucl. Phys. B485 (1997) 85, {\tt hep-th/9608024}.
\bibitem{DWPP} B. de Wit, K. Peeters and J.C. Plefka, {\it 
Supermembranes with winding}, to appear in Phys. Lett. B, 
{\tt hep-th/9705225} 
\bm{dWMN} B.\ de Wit, U.\ Marquard and H.\ Nicolai, \CMP{128}{1990}{39}.
\bibitem{Mtheory} P.K.\ Townsend, {\it Four Lectures on M-theory}, 
lectures given at the 1996 ICTP Summer School in High Energy 
Physics and Cosmology, Trieste, {\tt hep-th/9612121}.
\bibitem{Polchinski} J.\ Polchinski, Phys.\ Rev.\ Lett.\ 75 (1995) 
4724, {\tt hep-th/9510017}.
\bibitem{Townsend} P.K.\ Townsend, Phys.\ Lett.\ B350 (1995) 184,
{\tt hep-th/9501068}, Phys.\ Lett.\ B373 (1996) 184, {\tt hep-th/9512062}.
\bibitem{witten3} E.\ Witten, Nucl.\ Phys.\ B443 (1995) 85, {\tt
hep-th/9503124}. 
\bm{dWLN} B. de Wit, M. L\"uscher and H. Nicolai,
\NPB{320}{1989}{135}.
\bm{BFSS} T. Banks, W. Fischler, S.H. Shenker and L. Susskind, 
\PRD{55}{1997}{5112}, {\tt hep-th/9610043}.
\bibitem{Tdual} W. Taylor, \PLB{394}{1997}{283}, {\tt hep-th/9611042};\\
O.Ganor, S.Rangoolam and W.Taylor, Nucl. Phys. B492 (1997) 191, 
{\tt hep-th/9611202};\\
L.Susskind,~{\it T Duality in M(atrix) Theory and S 
Duality in Field Theory}, 
{\tt hep-th/9611164}.
\bm{BSS} T. Banks, N. Seiberg and S.H. Shenker, {\it Branes from Matrices},
{\tt hep-th/9612157}.
\bibitem{BBPT} K. Becker and M. Becker, {\it A two-loop test of 
M(atrix) theory}, {\tt hep-th/9705091}\\
K. Becker, M. Becker, J. Polchinski and A. Tseytlin, {\it Higher 
order graviton scattering in M(atrix) theory}, {\tt 
hep-th/9706072}
\bibitem{matrixstrings}
L. Motl, {\it Proposals on Nonperturbative Superstring Interactions},  
{\tt hep-th/9701025} ;\\
T. Banks and N. Seiberg, {\it Strings from Matrices}, {\tt hep-th/9702187};\\
S. Sethi and L. Susskind, {\it Rotational Invariance in the M(atrix) 
Formulation of Type IIB Theory}, {\tt hep-th/9702101}; \\
R. Dijkgraaf, E. Verlinde and H. Verlinde, {\it Matrix String Theory}, 
{\tt hep-th/9703030}.
\bibitem{EMM} K.~Ezawa,~Y.~Matsue~and~K.~Murakami, {\it Lo\-rentz Symmetry 
of Supermembrane in Light Cone Gauge Formulation}, {\tt hep-th/9705005}.
\bibitem{DWN} B. de Wit and H. Nicolai, in proc. Trieste Conference on 
Supermembranes and Physics in $2+1$ dimensions, p. 196, eds. M.J. Duff, 
C.N. Pope and E. Sezgin (World Scient., 1990);\\ 
B.\ de Wit, Nucl. Phys. B (Proc. 
Suppl.) 56B (1997) 76, {\tt hep-th/97101169}.
\bibitem{HF} J. Fr\"ohlich and J. Hoppe, {\it On zero-mass ground 
states in super-membrane matrix models},  {\tt hep-th/9701119}.
\bibitem{SS} S. Sethi and M. Stern, {\it D-Brane Bound States 
Redux}, {\tt hep-th/9705046}. 
\bm{Russo} K. Fujikawa and J. Kubo, \NPB{356}{1991}{208};\\
J.G. Russo, \NPB{492}{1997}{205}, {\tt hep-th/9610018};\\
R. Kallosh, {\it Wrapped Supermembranes}, {\tt hep-th/9612004}.
\bibitem{recent} K. Ezawa,~Y. Matsue and K. Murakami, {\it BPS 
Configuration of Supermembrane with Winding in M-direction}, {\tt 
hep-th/9706002}; \\
K. Fujikawa and K. Okuyama, {\it On a Lorentz 
covariant matrix regularization of membrane theories}, {\tt 
hep-th/9706027};\\  
I. Martin, A. Restuccia and R. Torrealba, {\it On the 
stability of compactified $D=11$ supermem\-branes}, {\tt 
hep-th/9706090}.
\end{thebibliography}
\end{document}